\documentstyle[12pt]{article}

\def\pa{\partial}

\def\a{\alpha}

\def\d{\delta}

\def\e{\epsilon}

\def\k{\kappa}

\def\m{\mu}
\def\n{\nu}

\def\mn{{\mu\nu}}
\def\ab{{\alpha\beta}}
\def\be{\begin{equation}}
\def\bea{\begin{eqnarray}}
\def\ee{\end{equation}}
\def\eea{\end{eqnarray}}
\def\rv{{\bf r}}
\def\jv{{\bf j}}
\def\kv{{\bf k}}
\def\gradv{\mbox{\boldmath$\nabla$}}
\def\av{{\bf A}}
\def\bv{{\bf B}}
\def\ev{{\bf E}}

\begin{document}

\hfill BRX TH-549

\begin{center}
{\bf How Special Relativity Determines the Signs of \\the
Nonrelativistic Coulomb and Newtonian Forces}

S.\ Deser\footnote{\tt email: deser@brandeis.edu}\\
Department of Physics, Brandeis University\\
Waltham, Massachusetts 02454
\end{center}

\abstract{We show that the empirical signs of the fundamental {\it
static} Coulomb/Newton forces are dictated by the seemingly
unrelated requirement that the photons/gravitons in the respective
underlying Maxwell/Einstein physics be stable. This linkage, which
is imposed by special relativity, is manifested upon decomposing
the corresponding fields and sources in a gauge-invariant way, and
without appeal to static limits. The signs of these free field
excitation energies determine those of the instantaneous forces
between sources; opposite Coulomb/Newton signs are direct
consequences of the Maxwell/Einstein free excitations' odd/even
spins.}

\section{Introduction}

One of the less heralded triumphs of special relativity (SR) is
that it determines the signs of the interactions between sources
according to the spins of their mediating fields. In contrast,
these signs are arbitrary in non-relativistic physics: the
observed Coulomb/Newtonian repulsion/attraction must be put in by
hand. SR bans instantaneous action-at-a-distance in favor of
mediating, and necessarily dynamical, fields. The resulting
Maxwell/Einstein framework then predicts these static properties.
As we will see, the (lightlike) free excitations' energy signs are
rigidly (if not obviously) linked to those of the nonrelativistic
regime's source-source interactions, where these (classical)
``photons" and ``gravitons" otherwise play no role at all. The
interactions' signs are determined by the odd/even spins of the
mediating fields.

We will carry out the derivations both by a simple static limit
shortcut and by a (more elaborate) gauge-invariant procedure,
where time-independence is not invoked. One brief Appendix extends
our results to the more general, but less physical, systems of
arbitrary spin and to form fields; the second provides a quick
covariant (but more technical) exposition of the phenomenon.

\section{Mediating Fields}

Non-relativistically, action-at-a-distance is translated into a
local field framework by defining a scalar potential field $\phi$
with the action
 \be I_{\rm nr}[\phi ;\rho ] = {\e \over 2}\!
\int\! d^dx \phi (- \nabla^2) \phi + \int\! d^dx \rho \phi, \ee
 which is to be added to the free particle actions. Here, $\e =
\pm 1$ is a sign factor, $\rho$ is the particle density, and $d$
is the space dimensionality, which does not affect the analysis.
[We also could include a parameter $m^2$ to cover both infinite
($m=0$) and finite ($m\neq0$) range forces by using the (positive)
Yukawa operator $(-\nabla^2 +m^2)$.] The sign of the force between
particles is obtained after a field-redefinition, $\phi =
\tilde{\phi} + \e G \rho$, where $-G$ is the usual Coulomb Green
function,
 \be
  \nabla^2 G(\rv-\rv')
= \d^d (\rv-\rv') .
 \ee
  Equation~(1) is recast in terms of $G$ as
\be I_{\rm nr}[\tilde{\phi} ;\rho ] = {\e\over 2}\! \int\! d^dx
\tilde{\phi} (-\nabla^2) \tilde{\phi} +  {\e\over 2}\!\int\! d^dx
\rho G \rho .
 \ee
 The free $\tilde{\phi}$-field obeys the
Laplace equation and simply decouples; the net interaction resides
entirely in the second term of Eq.~(3), whose sign depends only on
that of the free-field action. Because this sign, $\e$, is
arbitrary, the choice of attraction/repulsion $(\e = \mp)$ has to
be inserted by hand. [To check this sign correlation, write $\rho
= q_1 \d^d (\rv-\rv_1) + q_2 \d^d (\rv-\rv_2)$, that is, as a sum
of point sources, keep the cross terms, remembering that a
positive potential term in an action, $\int$(T--V), corresponds to
attractive, negative, $V$.]

The first example of how SR determines everything is the
(non-gauge) scalar field itself. We must first promote the
Laplacian to the wave operator, $\nabla^2 \rightarrow \Box \equiv
\nabla^2 - \pa^2/\pa (ct)^2$; the free field part of the action
(1) then becomes ($c=1$ henceforth), after an integration by
parts,
 \be I_s = {\textstyle{\frac{1}{2}}}\! \int\! d^dx dt \big\{  \big[
\dot{\phi}^2 + \phi \nabla^2 \phi \big] + 2\rho \phi\big\} \; .
 \ee
 The relative sign in Eq.~(4) is thus fixed by SR; its overall sign
ensures that the scalar field's newly acquired free excitation
mode has positive energy with respect to the usual convention for
a free particle's, $I_p = \frac{m}{2} \!\int dt \dot{x}^2$.
Otherwise, there would be no stable ground state because as the
particles radiated the field away, they would {\it gain} energy!
So a scalar's $\e$ sign is {\it necessarily} negative,
corresponding to {\it attraction} between like sources. To
summarize, SR here forced the sign of the static,
``action-at-a-distance,'' part of the action by the seemingly
remote two-step requirements of ``covariantization",
$\nabla^2\rightarrow\Box$, and of positive kinetic energy of the
resulting free field excitations.

\section{Maxwell}

We now come to the first physical example, Coulomb repulsion. The
Maxwell field's action is
\bea
I_M [A_\m, j^\m ] & \equiv & - \frac{1}{4}\!\int\! dt
d^dx [F_\mn F^\mn  -4 A_\m j^\m ] \nonumber \\
& = & \!\int\! dt\, d^dx \Big[ \frac{1}{2} \big((\nabla A_0 -
\dot\av)^2- (\gradv \times \av)^2) + \jv \cdot \av + j^0A_0 \Big]
, \eea in terms of $F_\mn = \pa_\m A_\n - \pa_\n A_\m$, with
signature $(-,+++)$. The static, Coulomb, force concerns only
$j^0$, and does not involve the vector potential $\av$, although
$\av$ alone determines the overall sign of Eq.~(5) and thereby of
the force. That is, the sign of the action is again fixed by the
positivity of the kinetic term in $I_M$ which describes the pure
``photon" excitations, \be I_M = + \!\int\! dt\,d^dx \Big\{
{\textstyle{\frac{1}{2}}} \big[ \dot \av^2 - (\gradv \times \av)^2
\big] + {\textstyle{\frac{1}{2}}} ( \gradv A_0)^2 + j^0 A_0 + \jv
\cdot \av - \gradv A_0 \cdot \dot \av \Big\} . \ee Indeed, $A_0$
is {\it not} dynamical at all, but an auxiliary variable that
enforces Gauss's law. The time-independent, Coulomb, part of
Eq.~(6) is then
$$
I_M \rightarrow \!\int\! dt\, d^d x \big[ -
\textstyle{\frac{1}{2}} A_0 \nabla^2 A_0 + j^0 \: A_0 \big] .
\eqno(7{\rm a})
$$

We also may follow the scalar field's redefinition procedure used
to reach Eq.~(3): Let $A_0 \rightarrow \tilde{A}_0 + G (j^0 +
\gradv \cdot \dot \av)$, which again leads to a decoupled
$\tilde{A}_0$ field and the ``residual" repulsion
$$
I \rightarrow {\textstyle{\frac{1}{2}}}\! \int\! d^dx\, j^0 G j^0
. \eqno(7{\rm b})
$$

The above discussion has the drawback that it is not entirely
gauge invariant, and makes the implicit gauge choice $\dot \av^L =
0$ to obtain Eq.~(7); also the nonrelativistic limit is not
needed. Potentials can be eliminated altogether by using the
continuity equation, $\gradv \cdot \jv + \pa_0 \: j^0 = 0$ (forced
by gauge invariance) to remove the current's divergence in terms
of $j^0$, and hence to write the coupling as $\int\! j^0 {\cal
E}$, where the scalar ${\cal E}$ is essentially the divergence of
the electric field $\ev \equiv \gradv A_0 -\dot{\av}$, namely
${\cal E} \equiv G \gradv \cdot \ev$. For this purpose, we resort
to the orthogonal decomposition of any vector field,
$$
\av = \av^T + \av^L, \quad \gradv \cdot \av^T = 0 = \nabla \times
\av^L, \quad \int\! d^dx\, \av^T \cdot \bv^L = 0; \eqno(8{\rm a})
$$
the orthogonality between any two $T$ and $L$ vectors expressed in
the last equation is especially important. This partition of a
vector field is the Fourier transform of the simple momentum space
algebraic decomposition,
$$
\av(\kv) = - \hat\kv \times (\hat\kv \times \av) + \hat \kv (\hat
\kv \cdot \av ) \equiv \av^T( \kv ) + \av^L(\kv ), \quad \av^T(\kv
) \cdot \bv^L (\kv) = 0 , \eqno(8{\rm b})
$$
in terms of some unit vector $\hat{\kv}$ (using $d=3$ notation).
The part of the action (6) involving $j^0$ is the sum of the
coupling and kinetic Maxwell terms:
\renewcommand{\theequation}{\arabic{equation}}
\setcounter{equation}{8} \be I_M \rightarrow \int dt d^dx \left[ -
\textstyle{\frac{1}{2}} {\cal E} \nabla^2 {\cal E} + j^0 \: {\cal
E} \right] . \ee [We need not belabor the now familiar drill,
defining ${\cal E} \rightarrow \tilde{\cal E} + G \: j^0$, etc.]
Both Eqs.~(7) and (9) lead to Coulomb repulsion, as is clear
because the ``potentials" $A_0$ or ${\cal E}$ appear with opposite
$\e$-sign to the (attractive) scalar potentials. [Parenthetically,
another fundamental byproduct of SR is that Maxwell's (and
Einstein's) equations contain additional static information that
is unavailable to nonrelativistic descriptions: the time-constancy
of the electric charge and gravitational mass~\cite{Franklin}.
These conservation laws follows from the exclusion of monopole
radiation in gauge theories, whereas nothing forbids time-varying
``charges" non-relativistically or for scalar fields.] Finite
range vector fields merely differ from Maxwell's by the addition
of a term \be I_m(A) = \frac{m^2}{2} \!\int\! dt\,d^dx [A^2_0 -
\av^2], \ee resulting in the shift from the infinite range Coulomb
to a (still repulsive) Yukawa interaction. Even the sign of $m^2$
is fixed by physics: changing it results in tachyonic propagation
of the field excitations, and the relative sign between $A^2_0$
and $\av^2$ is forced by Lorentz covariance, $\av^2 - A^2_0 = A_\m
A^\m$, in terms of the 4-vector potential $A_\m$. Thus, even if
electrodynamics had a finite range, our sign conclusions would be
unaffected.

\section{Gravity}

We turn now to our other main subject, gravity. Reduction to the
Newtonian limit of full general relativity is rather complicated;
even the notion of static limit must be analyzed carefully,
because in this theory with space-time coordinate invariance,
``static" means with respect to an ``inertial" frame. Furthermore,
the Newtonian limit (see, for example, Ref.~\cite{ADM}) involves
weak slowly moving sources or large separations between heavy
ones. Nevertheless, the physical upshot is effectively that (after
these tricky safeguards are understood) the force is governed by
the weak gravity limit, namely the linear massless spin 2 field.
We therefore turn to the latter, starting with its action and
field equations in terms of the linearized Einstein tensor
$G^L_\mn$:
 \be
I_2 [h_\mn ;T^\mn ]  =  \!\int\! dt d^dx \left[ h_\mn G^\mn_L(h) +
\k T^\mn h_\mn \right],
 \ee
 \be 2 G^L_\mn  \equiv  \Box h_\mn
-(\pa^2_{\a\n} h^\a~\!\!_\m + \pa^2_{\a\m} h^\a~\!\!_\n ) +
\pa^2_\mn h^\a_\a - \eta_{\mn} (\Box h^\a_\a - \pa^2_{\ab}h^\ab) =
- \k T_\mn .
 \ee
 The overall sign of $I_2$ yields the
$+ \frac{1}{2} \!\int \dot{h}^2_{ij}$ leading graviton kinetic
term. By (12), the Bianchi identity, $\pa_\m G^\mn_L \equiv 0$,
forces conservation of $T^\mn$. [It is also easy to verify the
action's gauge invariance under $\d h_\mn = \pa_\m \xi_\n + \pa_\n
\xi_\m$, since also $G^\mn_L (h + \d h) \equiv G^\mn_L (h)$.]

Following first the static limit approach, only two fields,
$h_{00}$ and $\nabla^2 h^T \equiv \frac{1}{2} (\d_{ij} \nabla^2 -
\pa^2_{ij})h_{ij}$ are relevant; the former is the counterpart of
$A_0$, and the latter plays the role of ${\cal E}$ and is (like
${\cal E}$) gauge-invariant. In this static limit approach, which
(as for Maxwell) is a gauge-dependent procedure it is a
straightforward consequence of (11,12) that
 \be I_2 [h_{00} ; T_{00}]
\rightarrow \!\int\! dt d^dx \Big[ h_{00} (\k T_{00} + \nabla^2
h^T ) - \frac{1}{4} h^T \nabla^2 h^T \Big] \rightarrow
-\frac{1}{4} \k^2 \!\int\! dt d^dx T_{00}\: G \: T_{00} .
 \ee
  Here and
henceforth, we specialize to our $d=3$ world; the generic $d$
dependence is given in Appendix B. The first integral shows (in
the ``Coulomb'' gauge) the action's reduced field dependence for
weak static $T^{00}$; the second, its form upon eliminating the
``Newton" constraint $\nabla^2 h^T +\k T_{00} = 0$. Attraction
between like, that is, positive mass, particles follows
irrespective of the sign of $\k$. [See Ref.~\cite{DeserPirani} for
some amusing generalizations.] The magnitude of $\k^2$ is twice
that of the Newtonian constant, as defined nonrelativistically in
(3).

We now come to the more refined treatment, where gauge invariance
is maintained and no static limit is required. The linearized
action (14) is first expressed with space and time components (as
well as orthogonal components of $h_{0i}$) separated,
 \bea \lefteqn{ I_2 [ h_{ij}, h_{0i} \equiv N_i^T + N^L_i,
h_{00} \equiv
N; T^{ij},T^{0i}, T^{00}] \equiv \int dt d^dx h_\mn G^\mn_L=} \nonumber \\
&& \textstyle{\frac{1}{2}} \int dt d^dx \Big\{ \Big[ h_{ij} \Box
h_{ij} - h_{ii} \Box h_{jj} + 2N\Box h_{ii} - 2 N^T_i \nabla^2
N^T_i \nonumber \\
&& - 2 N h_{ij,ij} + 2 h_{ii} (h_{\ell m,\ell m} - 2
\dot{N}_{i,i}
+\ddot{N} ) - 4 h_{ij,j} \dot{N}_i + 2 (h_{ij,j})^2 \Big] \nonumber \\
&& \hspace{.4in}+ 2\k (h_{ij} T^{ij} + 2N_iT^{0i} + NT^{00})
\Big\} ; \eea commas denote partial derivatives. We first retrace
the static limit results, keeping only the dependence on the
relevant variables:
$$
I_L \Big[ h_{ij}, N, \pa_t = 0 \Big] \rightarrow
\frac{1}{2} \!\int\! dt d^dx \Big[ h_{ij} \nabla^2 h_{ij}
- h_{ii} \nabla^2 h_{jj} + 2 N\nabla^2 h_{ii} - 2 N h_{ij,ij}
\nonumber
$$
$$
+ 2 h_{ii} h_{\ell m, \ell m} + 2\k N T^{00} \Big] .
~~~~~~~~~~~~~~~~~~~~~~~~ \eqno(15)
$$
The part of (15) involving  $T^{00}$, $N$ and $\nabla^2h^T$
correctly reduces to (13),
 \setcounter{equation}{15} \be I_L [h^T,
N, T^{00}] \rightarrow \!\int\! dt d^dx \Big[ N(\k T^{00} +
\nabla^2 h^T) - \textstyle{\frac{1}{4}} h^T \nabla^2 h^T \Big] .
\ee
 However, although $\nabla^2 h^T$, being the component $G^L_{00}$
of the (gauge invariant) linear Einstein tensor $G^L_\mn$ is also
invariant,  this reduction process does involve gauge choices
through assuming various gauge components of the metric to be
time-independent.

We will now indicate how to bypass these assumptions as well as
time-independence itself. Before doing so, we mention that
something else has been (usefully) bypassed here and by the next
procedure. We are obtaining the two-particle interaction term
directly, thereby avoiding the apparent textbook paradox that a
slowly moving particle's geodesic equation $\ddot \rv \cong
\frac{1}{2} \gradv h_{00}$, whereas it is the gauge invariant
component $h^T$ that ought to be the Poisson equation potential
according to Eq.~(16). The equivalence of $h_{00} \equiv N$ and
$h^T$ can obviously only be valid in certain ``static" gauges
\cite{ADM}.

To formulate the relevant part of Eq.~(14) in terms of gauge
invariants only, we begin by noting that the use of stress tensor
conservation, $\pa_\m T^\mn = 0$ (the linearized approximation is
in any case valid only for prescribed, conserved, sources) enables
us to rewrite the interaction term as:
 \bea \k \int\!dt\,d^dx \; h_\mn
T^\mn & = & \k \int\! dt\,d^dx\,\psi T^{00}, \nonumber \\
 \quad
\nabla ^4 \psi & \equiv & \nabla^4 N - 2\nabla^2 \dot{N}_{i,i} +
\ddot{h}_{ij,ij} \equiv \nabla^2 R_{00} - G_{ij,ij} . \eea Because
$\psi$ is a combination of (intrinsically gauge-invariant)
curvature components, its gauge invariance is guaranteed. We now
look for the other terms in Eq.~(14) that depend on $N$ (or
$\psi$), that is, the combination $\psi \nabla^2 h^T$. Finally, we
find the remaining dependence of Eq.~(14) on $h^T$ which is the
covariantized version of the static, $\int h^T \nabla^2 h^T$,
combination of Eq.~(16), after setting $\nabla^2 \rightarrow
\Box$. So the relevant gauge invariant, but {\it non}-static, part
of Eq.~(14) reduces to
 \bea
 I_L [\psi, h^T, T^{00}] & =
& \!\int dt d^dx \left\{ \psi [\k T^{00} + \nabla^2 h^T] -
\textstyle{\frac{1}{4}} h^T \Box h^T
\right\}  \nonumber \\
& = & - \frac{\k^2}{2}\! \int \! dt d^dx \left\{ T^{00} G T^{00}
-T^{00} \: GG \: \pa_0^2  \: T^{00} \right\} ,
 \eea
upon using $\Box\equiv\nabla^2-\pa_0^2 \equiv G^{-1} -\pa_0^2$ and
eliminating the now-familiar constraint.

At first sight, Eq.~(18) would seem to embody a retarded version
of the Newtonian law, but in fact we can remove the retardation:
the $\int T^{00} G G \pa^2_0 T^{00}$ term can be converted into an
instantaneous momentum interaction, using conservation,
$\dot{T}^{00} + \pa_i T^{0i} = 0$, to remove the time derivatives.
Then $\int \dot{T}^{00} GG \dot{T}^{00} = \int \pa_i T^{0i} GG
\pa_j T^{0j} = - \int T^{0i}_L G T^{0i}_L$, where the vector
$T^{0i}_L$ is the longitudinal momentum density. Because this is a
tensor theory, there are now both $T^{00}-T^{00}$ and
$T^{0i}-T^{0i}$ instantaneous interactions. For slow particles,
$T^{0i}=0$, and only the Newtonian force survives.

We have provided a gauge and Lorentz invariant treatment of weak
gravity that yields (without taking explicit static limits)
precisely the instantaneous Newtonian force law between energy
densities. As in electrodynamics, manifest Lorentz invariance has
been given up for this privilege; Appendix B reassures us that it
is not really lost.

\section{Summary}

That Coulomb and Newtonian forces are subsumed in their
relativistic Maxwell and Einstein extensions is a truism. We have
tried to exhibit some of these theories' qualitative triumphs
based on this truism: The signs of their static, nonrelativistic
forces are not only fixed (and the total charges and masses
necessarily constant), but correlated to the (observationally
verified) stability of the fundamental, ultrarelativistic, free
field radiation, namely the (classical) photons and gravitons.
That is, we related the static forces' signs to those of the free
lightlike excitations that do not even couple to static sources:
Despite their qualitatively different roles, the static and
dynamic field components are linked kinematically by being part of
a single (vector or tensor) Lorentz entity and the corresponding
static force signs are correlated to the (odd or even) spins of
the fields.

\vspace{.2in}

\noindent{\Large\bf Appendix A: Forms and Higher Spins}

\vspace{.2in}

\renewcommand{\theequation}{A.\arabic{equation}}
\setcounter{equation}{0}

The Maxwell and Einstein actions have obvious extensions when we
attach more indices to the basic fields: they can enter
antisymmetrically -- the so-called form fields -- or symmetrically
as in gravity's 2-index metric field, not to mention fields of
mixed symmetry.

We begin with form fields, whose current interest is due to their
appearance in string theory. A form field has a totally
antisymmetric potential $A_{[\mn\ldots]}$, and associated field
strength $B_{[\l\mn\ldots]} = \pa_{[\l} A_{\mn\ldots ]}$ subject
to the action \be I_{\rm form} [A] = \!\int\! dt d^dx \left\{ +
{\textstyle{\frac{1}{2}}} \dot{A}^2_{[ij\ldots]} + \ldots +
J^{\mn\ldots}A_{\mn \ldots}\right\}, \ee which directly mimics
Maxwell's action but with an antisymmetric current $J^{[\mn
\ldots]}$ (square brackets denote total anti-symmetrization of
included indices). Clearly, the only departure from Maxwell lies
in the number of indices. Because there is still only one static
source $\sim J^{0i\ldots}$ coupled to $A_{0i\ldots}$, and the
spatial indices do not affect any signs upon being moved,
$J^{0i\ldots} = J^0\,_{i\ldots}$, we can conclude that like static
sources $J^{0i\ldots}$ repel each other, just as in the
``one-form," Maxwell case. (The one exception is the degenerate
``zero-form,'' that is, the scalar, where there are no indices at
all.)

The other main line, extension beyond symmetric 2-tensors is to
symmetric tensor fields, $h_{\mn\a\ldots}$. These systems describe
higher spin excitations, with spin values $s$, equal to the number
of indices of $h_{\mn\a\ldots}$. Here the essential
--- and to date only physical --- application is to (spin 2)
gravity. For all spins, the actions are of the form
 \be I_{s\geq
2} [h_{\mn \ldots};T^{\mn\ldots} ] = \frac{1}{2} \!\int\!
dt\,d^dx\Big\{ {\textstyle{\frac{1}{2}}}\: \dot{h}^2_{ij\ldots} +
\ldots\Big\} + \k \!\int\! dt\,d^dx T^{\mn\ldots} h_{\mn\ldots},
\ee
 where $T^{\mn\ldots}$ is (necessarily) a symmetric tensor. We
 have omitted the additional terms in
the free action required for gauge invariance, as well as mass
term that would appear in the finite range versions of (A2).
Actually,  spin $>$2 fields are prone to coupling inconsistencies,
have never been seen, and conserved dynamical (in contrast to
fixed) higher rank symmetric sources $T^{\mn\a\ldots}$ are
physically excluded \cite{Weinberg}. Apart from these little
problems, the alternation of signs of the force with spin follows
directly from Eq.~(12): The overall sign of the free action is
determined so that the propagating modes, $h_{ij\ldots}$, have
kinetic terms $+\frac{1}{2} \int(\dot{h}_{ij\ldots})^2$. This sign
again fixes that of the ``Newtonian" terms according to the number
of time indices involved: even/odd $s$ implies
attraction/repulsion, where $s$ simultaneously counts spin and
number of indices, by exactly the same analysis as for $s=2/1$ in
text.

\vspace{.2in}

\noindent{\Large\bf Appendix  B: A Covariant Derivation}

\vspace{.2in}

For the experts, we append a rapid covariant, but less detailed,
derivation of our results.  If one ``completes the squares" in the
covariant scalar (4), Maxwell (5) and Einstein (11) actions, using
the respective propagators (in any gauge, since the sources are
conserved), one obtains the standard expressions
 $$
 I_s [\rho ] = -\frac{1}{2} \int dt \: d^dx \: \rho \, \Box^{-1} \rho
 \rightarrow - \frac{1}{2} \int \rho G\rho \; , \eqno{({\rm B.1a})}
 $$
 $$
 I_{max} [j] = -\frac{1}{2} \int dt \: d^dx \: j^\m \Box^{-1} j_\m
 \rightarrow + \frac{1}{2} \int j^0 \, G \, j^0 \; , \eqno{({\rm B.1b})}
 $$
 $$
 I_2 [T] = - \frac{\k^2}{2} \int dt \: d^dx \: \left[ T^\mn \Box^{-1} T_\mn
 - (d-1)^{-1} \: T^\m_\m \Box^{-1}
 T^\n_\n \right]
 $$
 $$ \rightarrow
 - \frac{1}{2} \: \left( \frac{d-2}{d-1} \right) \k^2
 \int T^{00} G \, T^{00} \; , \eqno{({\rm B.1c})}
 $$
for the effective interactions and their static limits. Here
$\Box^{-1}$ is (say) the retarded propagator, whose static limit
is our $G$. The overall signs of all actions are identical, as
befits the fact that they come from the $\frac{1}{2} \int
\phi\Box\phi$, $ \frac{1}{2}\int A^\m \Box A_\m$ and $\frac{1}{2}
\int h_\mn \, \Box \, h^\mn$ kinetic terms, with the same sign to
ensure stable free excitations. Instead, the scalar/Coulomb/Newton
sign difference is entirely encoded in the last terms, according
to the number of zeros (equal to the number of the static source's
indices or spin) to be raised or lowered.

The novel term in the tensor case is due to the fact that the
graviton propagator involves a trace factor.  The special values
of $d$ arise as follows: There are no Newtonian forces in $d=2$
Einstein theory \cite{DesertHooft}, while at $d=1$ the Einstein
tensor vanishes identically so Eqs. (11,12) become inconsistent.

\medskip I thank F.\ Ravndal for insisting on the pedagogical interest of
this ancient lore (updated to include form fields), and J.\
Franklin for comments. This research was supported by NSF grant
PHY04-01667.

\end{document}